\documentclass[12pt]{article}
\title{ On the extra factor of two in the phase of neutrino
oscillations}
\author{ L.B.Okun\footnote{email: okun@heron.itep.ru},
M.G.Schepkin\footnote{email: schepkin@heron.itep.ru},
 I.S.Tsukerman\footnote{email: zuckerma@heron.itep.ru}\\
 State Research
Center\\Institute of Theoretical and Experimental Physics, \\
Moscow, 117218 Russia}
\date{}
\newcommand{\beq}{\begin{eqnarray}}
 \newcommand{\eeq}{\end{eqnarray}}
\newcommand{\be}{\begin{equation}}
 \newcommand{\ee}{\end{equation}}
\begin{document}
\maketitle

\begin{abstract}
Attempts to modify the standard expression for the phase in
neutrino oscillations by an extra factor of two are based on
misuse of quantum mechanics. Claims to present Bruno Pontecorvo
and his coauthors as  ``godfathers'' of this ``extra 2'' factor
are easily disproved by unbiased reading their articles.
\end{abstract}

As is well known, the standard phase $\phi$ for neutrino
oscillations is given by $$ \phi=\Delta  m^2 l/2E,$$ where in the
case of two neutrino flavours $\Delta m^2=m^2_1-m^2_2$, $m_1$ and
$m_2$ being the masses of two mass eigenstates, $E$ is the
neutrino energy, and $l$ is the  distance between the points of
neutrino creation and its absorption.

Recent publications \cite{1}-\cite{2} show that  extra factor of
two in the standard expression for $\phi$, which has been proposed
and advocated by a number of  authors  during last decade (see
corresponding references in \cite{1}-\cite{2} and extensive review
\cite{3}) is still under discussion. Moreover, authors of
\cite{1}-\cite{2} claim that, the ``godfathers'' of this notorious
``extra 2'' factor were Pontecorvo and his coauthors Gribov and
Bilenky \cite{4}-\cite{7}.  But  the ``extra 2'' appears in
\cite{1}-\cite{2} as a result of improper manipulations with the
dependence of neutrino wave-functions both on space and time
coordinates. This approach should be  confronted with that of
refs. \cite{4}-\cite{8}, where the dependence on time only was
considered. In such a simple approach ``extra 2'' can be either a
misprint or an arithmetical mistake. In \cite{4} the momenta of
two neutrinos were implicitly assumed to be equal to  one and the
same value $p$ (hence $\Delta p=0)$. Thus the origin of ``extra
2'' in \cite{4} is a simple arithmetical mistake. According to
S.M.Bilenky (private communication) its reappearance in \cite{5}
was caused by a misprint. This is confirmed by comparison with the
publication \cite{7} where this ``extra 2'' is absent. Note that
\cite{7} preceded other papers (\cite{8} and \cite{6}) in which
there was no ``extra 2''. Thus Pontecorvo is the author of the
standard expression for $\phi$.

It was argued later by Lipkin \cite{9} and Stodolsky \cite{10}
that physically it is proper to consider oscillations not in time,
the phase of which is proportional to $\Delta E\cdot t$, but in
space with phase proportional to $\Delta p\cdot l$, and to presume
$\Delta E=0$. It was argued in \cite{9,10} that in none of the
neutrino oscillation experiments   the time was measured, only
distances between creation and detection points. The absence of
clocks in these experiments
 allows to consider the behaviour of neutrinos as a stationary
 one.
 (In ref.{\cite{11} an  experiment with tagged neutrinos was
 mentioned as an example of measuring not only distance, but time
 as well. The tagging by an accompanying muon fixes the time when
 neutrino was created. Unfortunately this experiment at Serpukhov
 was terminated before giving physically meaningful results. Thus
 the arguments of \cite{9,10} have at present no  counterexamples.)

Conceptually the interference of two mass-eigenstate neutrinos
with $\Delta E=0$ and different momenta is similar to the
interference of the electrons passing through two slits. In the
latter case it is obvious that time plays no role, as the
experiment is ``clockless''. Absolutely the same refers to the
existing measurements of neutrino oscillations. It is important
that individual electrons interfere with themselves, though the
interference pattern from two slits is created by all electrons
which were observed in a given experiment. The time intervals
between different electrons are irrelevant. The same is true for
neutrinos. Time $t$ is not an observable  in the neutrino
oscillation experiments.

The advocates of ``extra 2'' are considering the neutrino
``clockless'' oscillation experiments by describing them both in
space and time on an equal footing. Then they assume that each
mass eigenstate gets in its rest frame a phase $m_i \tau_i$
($i=1,2$), where $\tau_1 \neq \tau_2$. Different values of proper
times $\tau_1$ and $\tau_2$ presume that interference takes place
between two wave-functions at two different points, which
contradicts the essence of quantum mechanics. If one allows the
interference in different points (in space, or time, or in
space-time) then one gets wrong results for the case of $\Delta E
= 0$, for $\Delta p =0$, as well as for the case when both,
$\Delta E$ and $\Delta p$ are simultaneously different from zero.
This leads to the extra factor 2 in the phase (see Appendix where
the case $\Delta E = 0$ is considered) and other wrong
conclusions. Thus one of them is that the scenario of equal
velocities of two mass eigenstates is preferred in ref.\cite{1} to
that of equal energies in spite of the fact that $ \Delta v=0$
scenario was shown \cite{11, 12} to contradict simple kinematics.
Another erroneous statement of \cite{1} is that in the decay
$\pi\to\mu\nu$, the $\nu$ denotes a mixture of $\nu_\mu$ and
$\nu_e$.

The  following legitimate question could be asked. If distance,
not time, is adequate for description of stationary oscillations,
why then for many years any discussion of kaon oscillations and
then of neutrino oscillations (see \cite{4}-\cite{8}) started with
time dependent phases? The answer lies in the naturalness of
applying the Schroedinger equation to massive particles (kaons)
at rest and of using their decay widths. However even with time at
the beginning of discussion its results were always expressed in
terms of distance $l$. Therefore it is better to consider neutrino
oscillations from the beginning in terms of distance.

Of course, the stationary picture in space is valid only for
discussing the simplest problem when only the propagation of
neutrinos is studied. The additional measurement of, say,
accompanying muon and the decaying pion call for consideration of
both space and time \cite{13}. We are  unaware of any consistent
description of such experiments in the framework
of quantum field theory. Using wave packets instead of
off-mass-shell propagators has its own subtleties. However none of
them can  justify the extra factor of two. \\

We thank M.Beuthe, H.Lipkin and P.Minkowski for fruitful correspondence.

The work was supported by the RFBR grant 00 - 15 - 96562.


\newpage

 {\bf Appendix}

\vspace{5mm}

Let us consider the case $\Delta E = 0$. \\

1) The phase for each mass eigenstate ($\nu_1$ or $\nu_2$) is:

$$ Et - pl = Et(1-vv_s)  = Et[(1-v)+(1-v_s)-(1-v)(1-v_s)] =
\frac{Et}{2} \left [\frac{1}{\gamma^2} + \frac{1}{\gamma_s^2}
\right ]~, $$ \\ where $v=p/E$, $v_s=l/t$ (index $s$ stands for
"spacial"), $l$ and $t$ are distances in space and time between
the points of creation and detection of neutrino. Term
$(1-v)(1-v_s)$ proportional to $1/(\gamma^2 \gamma_s^2)$ is
neglected because both $\gamma$ and $\gamma_s$ are huge. \\

2) Let us now consider both $\nu_1$ and $\nu_2$. If $l$ and $t$ are
the same for them, then $\gamma_s$ is also the same and it drops out
when we consider the phase difference:

$$  \phi = \phi_1 - \phi_2 =
\frac{Et}{2} \left [\frac{1}{\gamma_1^2} - \frac{1}{\gamma_2^2}
\right ] = \frac{(m_1^2-m_2^2)t}{2E}=\frac{(m_1^2-m_2^2)l}{2E}~. $$

At the final step we replaced $t$ by $l$ assuming that neutrinos
are ultrarelativistic and using $c$ as a unit of velocity.
Thus we arrive at the standard expression of $\phi$.  \\

3) One could derive an extra factor of two by assuming that each mass
eigenstate has its own $v_s$:

$$ v_s = v_1 = p_1/E_1 = 1- \frac{m_1^2}{2E}
\; \mbox{\rm ~~~for } \; \nu_1~, $$
$$v_s = v_2 = p_2/E_2 = 1- \frac{m_2^2}{2E}
\; \mbox{\rm ~~~for } \; \nu_2~, $$  \\

which is impossible for common values of $l$ and $t$ required by
quantum mechanics.

\end{document}